\documentclass[floatfix,twocolumn,aps,prl,reprint,superscriptaddress]{revtex4-1}
\pdfoutput=1

\usepackage{amsmath}
\usepackage{amssymb}
\usepackage{float}
\usepackage{graphicx}
\usepackage{color,soul}
\usepackage{hyperref}
\usepackage[colorinlistoftodos]{todonotes}
\usepackage{lipsum}

\graphicspath{ {figures/} {figures_supplement/} }

\makeatletter
\@ifundefined{textcolor}{}
{%
 \definecolor{BLACK}{gray}{0}
 \definecolor{WHITE}{gray}{1}
 \definecolor{RED}{rgb}{1,0,0}
 \definecolor{GREEN}{rgb}{0,1,0}
 \definecolor{BLUE}{rgb}{0,0,1}
 \definecolor{CYAN}{cmyk}{1,0,0,0}
 \definecolor{MAGENTA}{cmyk}{0,1,0,0}
 \definecolor{YELLOW}{cmyk}{0,0,1,0}
}
\usepackage{dcolumn}
\usepackage{bm}
\usepackage{float}
\makeatother

\begin{document}

\title{
Millimeter-Wave Four-Wave Mixing via Kinetic Inductance for Quantum Devices
}

\author{Alexander Anferov}
\affiliation{James Franck Institute, University of Chicago, Chicago, Illinois 60637, USA}
\affiliation{Department of Physics, University of Chicago, Chicago, Illinois 60637, USA}

\author{Aziza Suleymanzade}
\affiliation{James Franck Institute, University of Chicago, Chicago, Illinois 60637, USA}
\affiliation{Department of Physics, University of Chicago, Chicago, Illinois 60637, USA}

\author{Andrew Oriani}
\affiliation{Department of Physics, University of Chicago, Chicago, Illinois 60637, USA}
\affiliation{Pritzker School of Molecular Engineering, University of Chicago, Chicago, Illinois 60637, USA}

\author{Jonathan Simon}
\affiliation{James Franck Institute, University of Chicago, Chicago, Illinois 60637, USA}
\affiliation{Department of Physics, University of Chicago, Chicago, Illinois 60637, USA}
\affiliation{Pritzker School of Molecular Engineering, University of Chicago, Chicago, Illinois 60637, USA}

\author{David I. Schuster}
\email{David.Schuster@uchicago.edu}
\affiliation{James Franck Institute, University of Chicago, Chicago, Illinois 60637, USA}
\affiliation{Department of Physics, University of Chicago, Chicago, Illinois 60637, USA}
\affiliation{Pritzker School of Molecular Engineering, University of Chicago, Chicago, Illinois 60637, USA}

\date{\today }

\begin{abstract}
Millimeter-wave superconducting devices offer a platform for quantum experiments at temperatures above $1~$K, and new avenues for studying light-matter interactions in the strong coupling regime.
Using the intrinsic nonlinearity associated with kinetic inductance of thin film materials, we realize four-wave mixing at millimeter-wave frequencies, demonstrating a key component for superconducting quantum systems.
We report on the performance of niobium nitride resonators around $100~$GHz, patterned on thin ($20$-$50~$nm) films grown by atomic layer deposition, with sheet inductances up to $212~$pH/$\Box$ and critical temperatures up to $13.9~$K. For films thicker than $20~$nm, we measure quality factors from 1-6$ \times 10^4$, likely limited by two-level systems. 
Finally we measure degenerate parametric conversion for a $95~$GHz device with a forward efficiency up to $+16~$dB, paving the way for the development of nonlinear quantum devices at millimeter-wave frequencies.

\end{abstract}

\maketitle

For superconducting quantum circuits, the millimeter-wave spectrum presents a fascinating frequency regime between microwaves and optics, giving access to a wider range of energy scales, and lower sensitivity to thermal background noise due to higher photon energies. Many advances have been made refining microwave quantum devices \cite{devoret2013outlook,braginski2019superconductor}, typically relying on ultra-low temperatures in the millikelvin range to reduce sources of noise and quantum decoherence. Using millimeter-wave photons as building blocks for superconducting quantum devices offers transformative opportunities by allowing quantum experiments to be run at liquid Helium-4 temperatures,  allowing higher device power dissipation and enabling large scale direct integration with semiconductor devices \cite{braginski2019superconductor}. Millimeter-wave quantum devices could also provide new routes for studying strong-coupling light-matter interactions in this frequency regime \cite{xiang2013hybridrev,morton2011hybrid,vasilyev2004mmesr,aslam2015esr,raimond2001atomreview}, and present new opportunities for quantum-limited frequency conversion and detection \cite{pechal2017millimeter,tucker1985millimeterrev}.

Recent interest in next-generation communication devices \cite{niu2015surveyrev,bozzi2011review} has led to important advances in sensitive millimeter-wave measurement technology around $100~$GHz.
Realizing quantum systems at these frequencies however requires both the demonstration of low-loss components --- device materials with low absorption rates \cite{chang2014striploss,brown2016lossnb,zhang2012thzmetanbn} and resonators with long photon lifetimes \cite{hanham2003hiQ,kuhr2007ultrahiQ,shirokoff2012mkidhiQ,endo2013mmhiQ,gao2009mmhiQ,kongpop2016wbandhiQ} ---  and most importantly, elements providing nonlinear interactions, which for circuit quantum optics can be realized with four-wave mixing Kerr terms in the Hamiltonian. One approach commonly used at microwave frequencies relies on aluminum Josephson junctions \cite{braginski2019superconductor}, which yield necessary four-wave mixing at low powers. However to avoid breaking Cooper pairs with high-frequency photons, devices at millimeter-wave frequencies are limited to materials with higher superconducting critical temperatures ($T_c$). Higher $T_c$ junctions have been implemented as high-frequency mixers for millimeter-wave detection \cite{tucker1985millimeterrev,kerr2013mixer,mears1990mixer}, and ongoing efforts are improving losses for quantum applications \cite{grimm2017nbnJunction,olaya2019planarized}.

Kinetic inductance (KI) offers a promising alternative source of Kerr nonlinearity arising from the inertia of Cooper pairs in a superconductor, gaining recent interest for microwave quantum applications \cite{shearrow2018ald,samkharadze2016}, and has also been successfully used for millimeter-wave detection \cite{hailey2014mkid,noroozian2015mkid}. Niobium Nitride (NbN) is an ideal material for KI, as it has a high intrinsic sheet inductance \cite{ivry2014universalnbn,kamlapure2010nbngap,beebe2016nbn}, a relatively high $T_c$ between $14$-$18~$K \cite{ivry2014universalnbn,kamlapure2010nbngap,beebe2016nbn,sowa2017peald} making it suitable for high-frequency applications \cite{zhang2012thzmetanbn}, and has good microwave loss properties \cite{niepce2019nbnnanowirevap}. Among deposition methods, atomic layer deposition (ALD) offers conformal growth of NbN \cite{sowa2017peald} and promising avenues for realizing repeatable high KI devices on a wafer-scale \cite{shearrow2018ald}.

In this work, we explore kinetic inductance as a nonlinear element for quantum devices at millimeter-wave frequencies using high KI resonators in the W-Band (75-110GHz) fabricated from thin films of NbN deposited via ALD. We describe a method for characterizing resonances at single photon occupations, and study potential loss mechanisms at a wide range of powers. Using the power dependent frequency shift, we study the nonlinearity arising from KI, the strength of which varies with wire width and material properties. With two tone spectroscopy, we observe degenerate four-wave mixing near single photon powers. These measurements demonstrate the necessary core components for millimeter-wave circuit quantum optics, paving the way for a new generation of high-frequency high-temperature experiments.

\begin{figure}[t]
\centering
\includegraphics[width=3.37in]{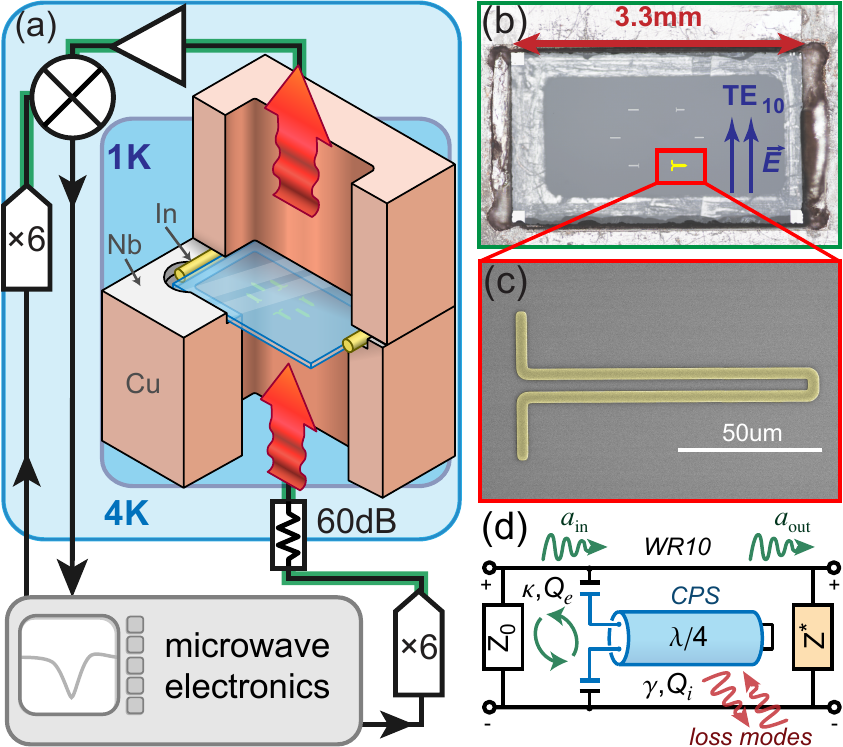}
\caption{Device characterization and design.
(a) Six-fold frequency multipliers convert microwave to millimeter-wave signals (green), which are demodulated with a cryogenic mixer. A cutaway shows copper WR-10 rectangular waveguides coupling the signal in and out of a Nb coated slot, into which we mount a chip patterned with 6 resonators. (b) Top down composite micrograph showing a mounted chip with the top waveguide removed. (c) Scanning Electron Micrograph of a typical resonator used in this work, with wire width $w=4~\mu$m and film thickness $t = 27.8~$nm (NbN false colored yellow). The dipole coupling antennas extend on the left side of the quarter-wave resonator. Measurements can be described with input and loss couplings $Q_e$ and $Q_i$ using the circuit model in (d), which takes into account the impedance mismatch between waveguide $Z_0$ and slot with sapphire chip $ Z^*$.
\label{fig:fig1}}
\end{figure}

We investigate properties of millimeter-wave high KI resonators in the quantum regime at temperatures of $1~$K in a Helium-4 adsorption refrigerator. Using a frequency multiplier, cryogenic mixer and low noise amplifier, we measure the complex transmission response as shown in Fig. \ref{fig:fig1}(a). Input attenuation reduces thermal noise reaching the sample, enabling transmission measurements in the single photon limit set by the thermal background. Rectangular waveguides couple the signal in and out of a $200~\mu$m deep slot, the dimensions of which are carefully selected to shift spurious lossy resonances out of the W-band. To reduce potential conductivity losses, the waveguide and slot are coated with $200~$nm of evaporated Niobium. Below $9~$K this helps shield the sample from stray magnetic fields, however devices with higher $T_c$ are not shielded from magnetic fields while cooling through their superconducting transition. We use indium to mount a chip patterned with 6 resonators in the slot, as shown in Fig. \ref{fig:fig1}(b). Devices are patterned on $100~\mu$m thick sapphire which has low dielectric loss, and minimizes spurious substrate resonances in the frequency band of interest. The planar resonator geometry shown in Fig. \ref{fig:fig1}(c) consists of a shorted quarter wave section of a balanced mode coplanar stripline waveguide (CPS), which couples directly to the TE$_{10}$ waveguide through dipole radiation, which we enhance with dipole antennas. We find that this design is well described by the analytic model presented in Ref. \cite{yoshida1992modeling} which takes into account the thin film linear kinetic inductance. For very thin or narrow wire widths, the total inductance is dominantly kinetic, making the resonators extremely sensitive to superconducting film properties. 

\begin{figure*}
\centering
\includegraphics[width=6.69in]{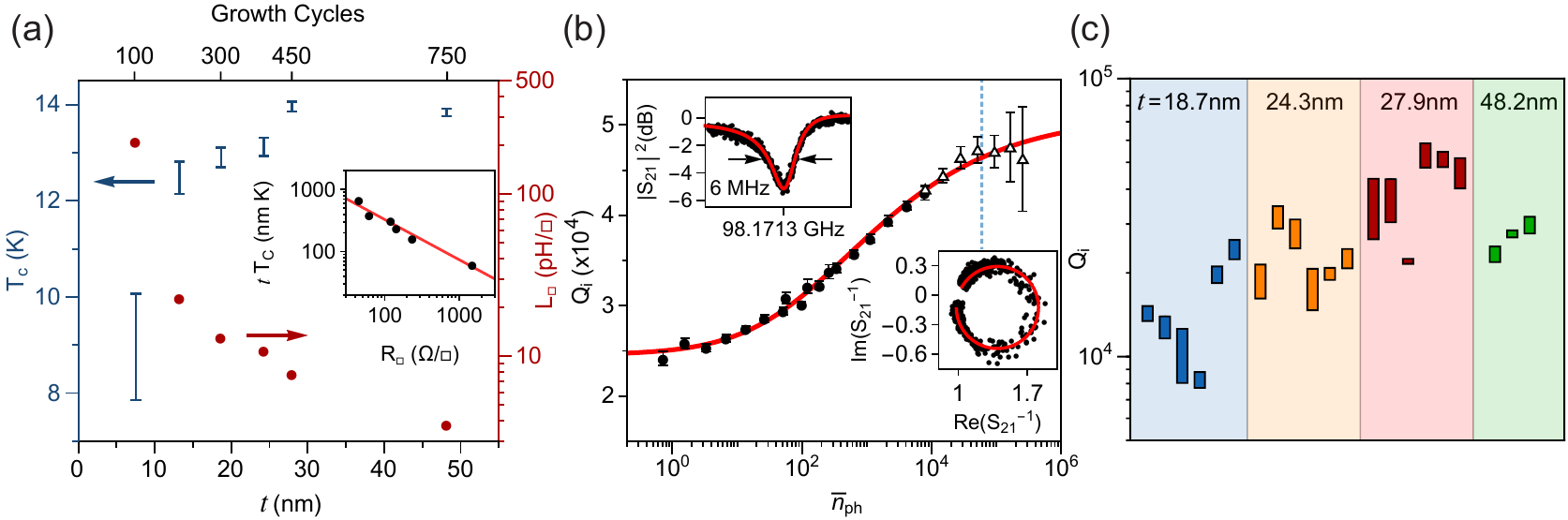}
\caption{
Material and loss characterization. (a) Superconducting critical temperature ($T_c$) and sheet inductance ($L_\Box$) of deposited NbN as a function of film thickness ($t$). Bars denote temperatures corresponding to 90\% and 10\% reductions from maximum resistivity. The inset shows the dependence of $tT_c$ on $R_\Box$ with a fit (red) to $tT_c = A R_\Box^{-B}$; $A=6487\pm1607$,$B=0.647\pm0.05$
(b) Power dependence of the internal quality factor for a resonator with $Q_e^* = 2.901 \times 10^4$ patterned on a $27.9$nm thick NbN film, measured at $1~$K. White triangles are fits to a nonlinear response model near and above the bifurcation power (dashed line), and the red line is a fit to a two-level system model including low and high power saturation. Insets show lineshape and fits at average photon occupation $\bar{n}_{\text{ph}} \approx 0.8$.
(c) Internal quality factors for resonators in this study, grouped by film thickness. The top and bottom of the colored bars correspond to fitted low and high power saturation values.
\label{fig:fig2}}
\end{figure*}

In order to understand the quality of the NbN films grown by ALD and accurately predict resonant frequencies, we characterize material properties with DC electrical measurements. All devices in this work are deposited on sapphire with a process based on Ref. \cite{sowa2017peald}, and etched with a fluorine based inductively coupled plasma.
We measure resistivity at ambient magnetic fields as a function of temperature, which we use to extract $T_c$ for a range of film thicknesses [See Fig. \ref{fig:fig2}(a)]. The inset also shows that our films follow a universal relation observed for thin superconducting films \cite{ivry2014universalnbn} linking $T_c$, film thickness $t$, and sheet resistance  $R_\Box$: we find that our results are similar to NbN deposited with other methods \cite{beebe2016nbn,ivry2014universalnbn}. For thicker films, $T_c$ appears to saturate at $13.8$-$13.9~$K which is comparable to other materials studies \cite{sowa2017peald,kamlapure2010nbngap,ivry2014universalnbn}, while decreasing to $8.7~$K for the thinnest film (t=$7~$nm), which can be attributed to disorder enhanced Coulomb repulsions \cite{skvortsov2005superconductivity,driessen2012strongly}. We also find that the superconducting transition width increases significantly for the thinner films, which can in turn be attributed to disorder broadened density of states \cite{driessen2012strongly} or reduced vortex-antivortex pairing energies at the transition \cite{niepce2019nbnnanowirevap,mooij1984percolationvap}.

From the resistivity and critical temperature we determine the sheet inductance $L_\Box=\hbar R_\Box/\pi \Delta _0$ where the normal sheet resistance $R_\Box = \rho_n/t$ is taken as the maximum value of normal resistivity $\rho_n$, occurring just above $T_c$, and $\Delta_0=2.08T_c$ is the superconducting energy gap predicted by BCS theory for NbN \cite{niepce2019nbnnanowirevap,kamlapure2010nbngap}. We observe a monotonic increase in $L_\Box$ for thinner films, achieving a maximum $L_\Box=212~$pH/$\Box$, comparable to similar high KI films \cite{shearrow2018ald,mondal2011phasenbn}.

By characterizing complex transmission spectra of resonators fabricated on a range of film thicknesses, we explore loss mechanisms at millimeter-wave frequencies. The sheet inductance, thickness, and $T_c$ measured for a given film are used to adjust the resonator design length. This spreads resonances out in frequency from $80~$GHz to $110~$GHz, while varying antenna lengths allows us to adjust coupling strengths. A typical normalized transmission spectrum taken at single photon occupation powers ($\bar{n}_{\text{ph}}\approx0.8$) is shown in the inset of Fig. \ref{fig:fig2}(b). On resonance, we observe a dip in magnitude, which at low powers is described well by \cite{khalil2012analysis}:
\begin{equation}
    S_{21}=1-\frac{Q}{Q_e^*}\frac{e^{i\phi}}{1 + 2i Q \frac{\omega-\omega_0}{\omega_0}}
    \label{eq:lowpowerS}
\end{equation}where $Q^{-1}=Q_i^{-1}+\text{Re}[Q_e^{-1}]$ \cite{khalil2012analysis} and the coupling quality factor $Q_e= Q_e^* e^{-i\phi}$ has undergone a complex rotation $\phi$ due to an impedance mismatch \cite{khalil2012analysis,megrant2012cleland}, likely induced by the sapphire chip and slot altering waveguide geometry. We find that at high powers, $Q_i$ begins to saturate, typically for $\bar{n}_{\text{ph}} > 10^5$ [Fig. \ref{fig:fig2}(b)]. This behavior is well described by a power dependent saturation mechanism \cite{sage2011studytls,wang2009improvingtls}, likely originating from two-level systems in the slow growing amorphous surface oxide layer \cite{medeiros2019measuringnbn}. For some samples, due to relatively low bifurcation powers we do not observe high power saturation. 

\begin{figure}[b]
\centering
\includegraphics[width=3.37in]{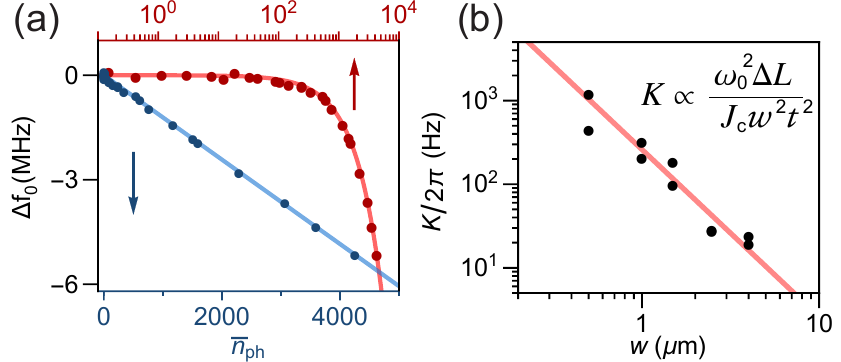}
\caption{Measuring Kerr nonlinearity (a) Frequency shift versus average photon number $\bar{n}_\text{ph}$ in the resonator in linear and log-scale respectively. (b) Extracted self-Kerr coefficients (accurate to within a factor of $\sim 10$) versus wire width $w$ for resonators fabricated from a $29~$nm thick film, with expected fit to $w^{-2}$ in red. We find no significant impact of $w$ on $Q_i$.
\label{fig:fig3}}
\end{figure}

\begin{figure*}
\centering
\includegraphics[width=6.69in]{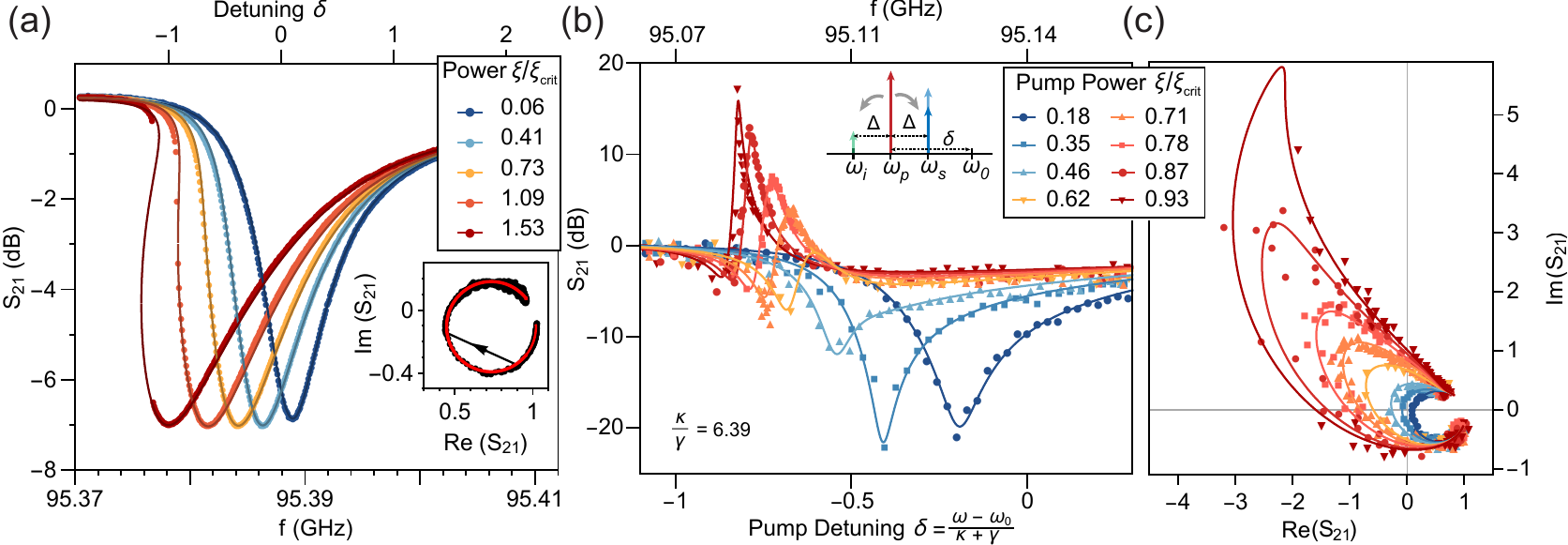}
\caption{
Nonlinear response and four-wave mixing. (a) Transmission of a typical resonance at a range of powers near and above bifurcation showing good agreement with a Kerr nonlinear response. The inset shows overlaid data and fits in the complex plane just below and above the bifurcation point. (b-c) Parametric conversion gain with a $95.1~$GHz device with the same film thickness as (a) as a function of reduced pump frequency $\delta$ for a fixed signal detuning $\Delta$ of +$450~$kHz, taken at increasing pump powers. Solid lines correspond to theoretical response. The initial forward de-amplification is better understood when the response is viewed in the complex plane (c), where we observe smooth parametric deformation from the single tone response.
\label{fig:fig4}}
\end{figure*}

To study the effect of film thickness on $Q_i$, we repeat the measurements summarized in Fig. \ref{fig:fig2}(b) for devices varying in thickness from 19.5 nm to 48.8 nm, and show the results in Fig. \ref{fig:fig2}(c), where we plot the low and high power saturation values of $Q_i$ for devices from six separate chips grouped by film thickness. For films thicker than 20nm, we consistently find $Q_i > 10^4$, all of which show power dependence to varying extents. We find a weak correlation of $Q_i$ with film thickness, which could be explained by several additional potential sources of loss.
Thinner films exhibiting higher disorder have also been connected with a nonlinear resistance associated with kinetic inductance \cite{yurke2006performance}.  Additionally, since the devices are not shielded from ambient magnetic fields at the superconducting transition, resulting vortices trapped in the thin films may lead to additional dissipation \cite{nsanzineza2016vortices,kitaygorsky2007darkvap,mooij1984percolationvap,niepce2019nbnnanowirevap} dependent on film thickness. Resonances patterned from thinner films proved experimentally difficult to distinguish from background fluctuations, possibly indicating low values of $Q_i$ or frequencies shifted out of the measurement bandwidth.

A key aspect of high KI resonators is their fourth-order nonlinearity: an important component for realizing quantum devices, and similar to the nonlinearity term found in Josephson junctions for low powers. 
Nonlinear kinetic inductance takes the general form $L= L_k + \Delta L_k I^2/I_c^2$, where $L_k$ is the linear kinetic inductance, $\Delta L_k$ the nonlinear kinetic inductance, and $I_c$ the critical current which sets the nonlinearity scale \cite{yurke2006performance,kher2017superconducting}. 
This adds nonlinear terms of the form $\frac{\hbar}{2} K(a^\dagger a)^2$ to the Hamiltonian, with $K\propto \omega_0^2 \Delta L_k /I_c^2$, shifting the fundamental frequency $\omega_0$ by the self-Kerr constant $K$ (or anharmonicity) for each photon added. To characterize the strength of the resonator nonlinearity, we measure the resonance frequency shift \footnote{The frequency corresponding to maximum photon occupation, or the point diametrically opposite $S_{21}=1$} 
as a function of photon number \cite{maleeva2018circuit}. A linear fit for a resonator ($t=29~$nm, $w=0.5~\mu$m)  yielding $K/2\pi=1.21^{+1.8}_{-0.8}~$kHz is shown in Fig. \ref{fig:fig3}(a). By writing the self-Kerr coefficient in terms of a current density $I_c = J_c w t$, we find that $K$ scales as $w^{-2}$, which we observe in \ref{fig:fig3}(b). These results are comparable to self-Kerr strengths of granular aluminum nanowires \cite{maleeva2018circuit} or weakly nonlinear Josephson junctions \cite{eichler2014controlling}. Despite careful calibration of input powers supplied to the device, we note that minute variations in received power arising from shifting attenuation at cryogenic temperatures limit best estimates of photon number to within a factor of $\sim$10. 

A hallmark of Kerr nonlinearity is the distortion of the transmission line-shape in frequency space at high powers, ultimately leading to a multi-valued response above the bifurcation power. Re-writing $\gamma = \omega_0 /Q_i$ and $\kappa = \omega_0 /\text{Re}[{Q}_e]$, the steady-state nonlinear response takes the form derived from Refs.~\cite{yurke2006performance,swenson2013operation} (See Appendix A):
\begin{equation}\label{eq:snonlin}
    S_{21} = 1 - \frac{\kappa}{\kappa + \gamma}
    \frac{1+ i \tan \phi}{1+2 i(\delta - \xi n)}
\end{equation}
where the frequency detuning is written in reduced form $\delta=\frac{\omega-\omega_0}{\kappa + \gamma}$, and $n=n_{\text{ph}}/\tilde{n}_{\text{in}}$ is a function of frequency and reduced circulating power $\xi = K\frac{\kappa}{hf(\kappa + \gamma)^3}P_\text{in}$. We plot steady-state transmission data taken near the bifurcation power in Fig. \ref{fig:fig4}(a) along with fits to Eq. \ref{eq:snonlin}, with system parameters $\kappa$, $\phi$  and $\omega_0$ constrained to low power values, and find the model in good agreement with measurements. 

We further explore nonlinear dynamics by stimulating degenerate four-wave mixing with the addition of a continuous wave classical pump (see supplement Fig. S3). When a high power pump tone is on resonance with the down-shifted resonance frequency, and a low power signal is at a frequency detuning $\Delta$ from the pump, we expect to observe the net conversion of two pump photons into a signal photon and an idler photon at their sum-average frequency [Fig. \ref{fig:fig4}(b) inset]. This effect is most pronounced when all frequencies are within the resonant bandwidth, and the pump power $\xi$ approaches the bifurcation point $\xi_\text{crit}$, but is limited by the loss fraction $\gamma/\kappa$. We measure the pump-signal conversion efficiency of a high-bandwidth, high-$K$ device in the propagation direction as a function of reduced pump frequency $\delta$ for increasing pump powers $\xi$, and a fixed signal power corresponding to $\bar{n}_{\text{ph}} \simeq 9$ in Fig. \ref{fig:fig4}(b-c). We find that this behavior is accurately captured with a model based on Refs. \cite{eichler2014controlling,yurke2006performance} (see Appendix A), and overlay the results. 
For increasing pump powers, we observe smooth parametric deformation from the single tone response in the complex plane. For higher powers, we observe increasing gain with decreasing linewidth similar to Refs. \cite{eichler2014controlling,tholen2007nonlinearities}, up to a maximum measured forward efficiency of +$16~$dB.
The slight curvature in the complex plane is a result of the finite pump-signal detuning $\Delta$.

The demonstration of degenerate four-wave mixing realizes an important milestone for the development of quantum devices at millimeter-wave frequencies and temperatures above $1~$K. For NbN films thicker than $25~$nm, we measured millimeter-wave resonators with internal quality factors exceeding $2\times 10^4$ at single photon powers, and by reducing wire width to $500~$nm achieved self-Kerr nonlinearities up to $1.21~$kHz for linewidths ranging from $1$-$200~$MHz. With some modification the devices in this work could easily be redesigned as parametric amplifiers, which at microwave frequencies have been shown to achieve near quantum-limited noise figures and quadrature squeezing \cite{tholen2007nonlinearities,eom2012widebandjpa,tholen2007nonlinearities,yurke1987squeezed,movshovich1990squeezing}. While insufficient for implementing a millimeter-wave artificial atom, the Kerr nonlinearity we measure arising from high KI thin films can further be used for superinductors \cite{bell2012superinductor,niepce2019nbnnanowirevap}, photon frequency conversion \cite{pechal2017millimeter}, parametric mode cooling \cite{khan2015crossK,zhang2017crossK}, phase slip junctions \cite{mooij2006phaseslip,astafiev2012coherentphaseslip}, and mode squeezing \cite{tholen2007nonlinearities} realized at millimeter-wave frequencies. This opens the door to a new generation of high-frequency quantum experiments at temperatures above $1~$K.

\begin{acknowledgments} 
 The authors would like to thank P. Duda, P. S. Barry, and E. Shirokoff for assistance developing deposition recipes, as well as M. Wesson and C. Sheagren for supporting film characterization. We acknowledge useful discussions with S. J. Whiteley, and also thank J. Jureller for assistance with MRSEC facilities. This work was supported by the Army Research Office under Grant No. W911NF-17-C-0024. This work was partially supported by the University of Chicago Materials Research Science and Engineering Center, which is funded by the National Science Foundation under award number DMR-1420709. This work was partially supported by the National Science Foundation Graduate Research Fellowship under Grant No. DGE-1746045. Devices were fabricated in the Pritzker Nanofabrication Facility at the University of Chicago, which receives support from Soft and Hybrid Nanotechnology Experimental (SHyNE) Resource (NSF ECCS-1542205), a node of the National Science Foundation’s National Nanotechnology Coordinated Infrastructure. 
\end{acknowledgments} 

\bibliography{thebibliography}

\onecolumngrid
\newpage
\appendix
\renewcommand{\thefigure}{S\arabic{figure}}
\setcounter{figure}{0}

\setcounter{equation}{0}
\numberwithin{equation}{section}
\renewcommand\theequation{\thesection\arabic{equation}}

\section{Kerr nonlinear dynamics for a side-coupled resonator}
\begin{figure}[htb]
\centering
\includegraphics[width=6.69in]{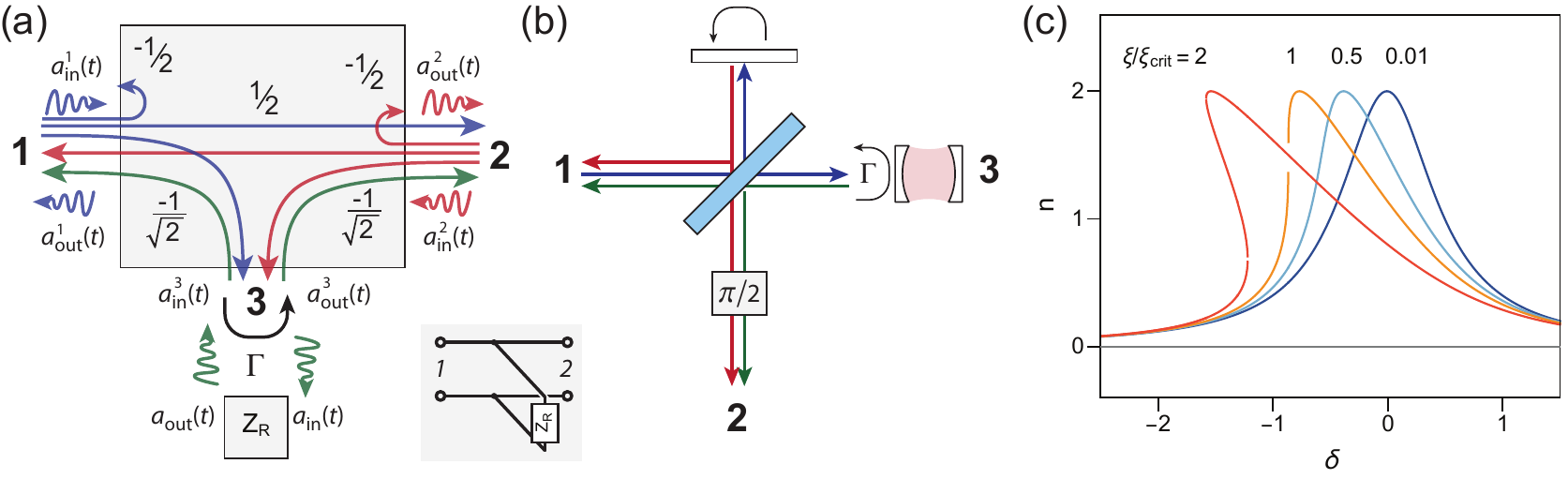}
\caption{Side-coupled resonator model (a) 3 port network of an H-plane splitter coupled to a black box resonator, showing corresponding transmission coefficients with input and output fields labelled by port. The inset shows the equivalent circuit network. (b) Analogous configuration for an optical cavity, adjusted for boundary conditions. (c) Normalized photon number as a function of reduced pump frequency $\delta$ for reduced drive strengths $\xi$, where $\xi_\text{crit}=-2/\sqrt{27}$.
\label{fig:figS1}}
\end{figure}

Here we outline a simple method inspired by Refs. \cite{mazin2005microwave,gao2008physics} to decompose a side-coupled resonator into a linear network containing a one-sided cavity, which is very well understood in the language of input-output theory used in quantum optics \cite{walls2007quantumoptics}. This allows us to map well-modelled dynamics of a Kerr nonlinear cavity driven in reflection \cite{yurke2006performance,eichler2014controlling} to a side-coupled resonator measured in transmission, obtaining results in agreement with Ref. \cite{swenson2013operation}, which uses a more direct approach.

Based on the circuit model in Fig. 1(d), and the argument that a symmetrically coupled resonator will radiate equally in both directions, we consider the the 3-port H-plane splitter. This lossless but unmatched network \cite{pozar2009microwave} has symmetric ports 1-2 corresponding to the transmission line, and unmatched port 3 leading to the single-port coupled resonator $Z_R$ as shown in Fig. \ref{fig:figS1}(a) (This system can also be described by a network consisting of a 50-50 beamsplitter, perfect mirror, and $\pi/2$ phase shifter as shown in Fig. \ref{fig:figS1}(b), which yields the same key results if we are careful to use correct boundary conditions). If we place a black-box element on port 3, we can describe it's input and output fields in terms of the waveguide input and output fields:
\begin{equation}
\left(\begin{array}{c}
a_{\text{in}}(t)\\
a_{\text{out}}(t)
\end{array}\right)=\frac{-1}{\sqrt{2}}
\left(\begin{array}{c}
a_1^{\text{in}}(t)+a_2^{\text{in}}(t)\\
a_1^{\text{out}}(t)+a_2^{\text{out}}(t)
\end{array}\right)
\label{Eq:inputinput}
\end{equation}
If we describe the black box with an arbitrary reflection term $\Gamma = a_{\text{out}}(t)/a_{\text{in}}(t)$, the scattering matrix of the system reduces to:
\begin{equation}
    S = 
\frac{1}{2}\left(
\begin{array}{cc}
\Gamma-1&\Gamma+1\\
\Gamma+1&\Gamma-1\\
\end{array}\right)
\label{Eq:s2matrix}
\end{equation}
We can now verify that far off-resonance, for open circuit perfect reflection $\Gamma\to1$, we recover perfect transmission. With a map of waveguide inputs and outputs we replace the black box with Kerr nonlinear one-port coupled resonator, which has the steady state condition \cite{walls2007quantumoptics,eichler2014controlling}:
\begin{equation}
    i(\omega-\omega_0)a + \frac{\kappa + \gamma}{2}a - i K |a|^2 a^* = \sqrt{\kappa}a_\text{in} = -\sqrt{\frac{\kappa}{2}}a_1^\text{in}
\label{Eq:steadystate}
\end{equation}
We have been careful to use the microwave convention for Fourier transforms, and $n_\text{ph}=|a|^2$ corresponds to the average number of photons in the resonator. Multiplying by the complex conjugate, we obtain an equation governing the normalized number of photons in the resonator $n$.
\begin{equation}
   \left(\frac{1}{4}+\delta^2\right) -2\delta \xi n^2 + \xi^2 n^3 =\frac{1}{2}
\label{Eq:nsolns}
\end{equation}
Where similar to Ref. \cite{eichler2014controlling}, $n$, $\xi$ and $\delta$ are defined as:
\begin{align}
    n&=\frac{|a|^2}{|a_1^\text{in}|^2}\frac{(\kappa+\gamma)^2}{\kappa}\\
    \xi &=\frac{|a_1^\text{in}|^2\kappa K}{(\kappa + \gamma)^3}\\
    \delta &= \frac{\omega-\omega_0}{\kappa + \gamma}
\end{align}
We plot $n$ as a function of $\delta$ for varying drive strengths $\xi$ in Fig. \ref{fig:figS1}(c), finding that $n$ reaches a maximum value of $2$. At the critical value $\xi_\text{crit}=-2/\sqrt{27}$, Eq. \ref{Eq:nsolns} has 3 real solutions, leading to the onset of bifurcation. 
Based on the resonator boundary conditions \cite{ridolfo2012photon} $a_\text{out} = a_{\text{in}}+a\sqrt{\kappa}$ and Eq. \ref{Eq:steadystate}, the reflection coefficient $\Gamma$ will be \cite{walls2007quantumoptics,yurke2006performance,eichler2014controlling}
\begin{equation}
    \Gamma =1-\frac{\kappa}{\kappa + \gamma}\frac{1}{\frac{1}{2}+i(\delta-\xi n)}
    \label{eq:gamma}
\end{equation}
Far off resonance, an impedance mismatch on output port 2 results in nonzero reflection $|r|=\sin \phi$ and transmission $|t|=\cos \phi$ less than unity. To account for this while preserving the unitarity of the S matrix, we apply transformations of the form $e^{i\phi}$ to each path of the 3 port network, yielding $S_{21} = (\Gamma e^{i\phi}+e^{-i\phi})/2$. Mapping Eq. \ref{eq:gamma} to the modified 3 port network, we obtain the result used in the main text, which  in respective limits agrees with Refs. \cite{khalil2012analysis} and \cite{swenson2013operation}.
\begin{equation}
    S_{21}=1-\frac{\kappa}{\kappa+\gamma}\frac{e^{i\phi}}{\cos \phi}\frac{1}{1+2i(\delta+\xi n)}
    \label{eq:Snonlin}
\end{equation}
At low powers ($\xi n\to0$), Eq. \ref{eq:Snonlin} reduces to Eq. 1 in the main text.

We follow a similar approach to obtain expressions for parametric conversion gain using the derived input-output relations to map the key results from Ref. \cite{eichler2014controlling} to the waveguide inputs and outputs. Using microwave conventions for Fourier transforms, the one-port gain of a signal detuned from the pump by $+\Delta = \frac{\omega_s-\omega_p}{\kappa + \gamma}$ is given by:
\begin{equation}
g_{s} = \frac{a_\Delta^\text{out}}{a_\Delta^\text{in}}=
1-\frac{\kappa }{\kappa + \gamma}\frac{\frac{1}{2}-i(\delta -2 \xi n-\Delta)}{(i\Delta +\lambda_+)(i \Delta + \lambda_-)}
\end{equation}
With $\lambda_\pm=\frac{1}{2}\pm\sqrt{(\xi n)^2-(\delta-2\xi n)^2}$. Using the 3 port network transformations above yields the normalized forward (in direction of propagation) signal gain:
\begin{equation}
g^+_{s} = \frac{a_{2,\Delta}^\text{out}}{a_{1,\Delta}^\text{in}}=
1-\frac{\kappa }{\kappa + \gamma}\frac{e^{i\phi}}{\cos \phi}\frac{\frac{1}{2}-i(\delta -2 \xi n-\Delta)}{2(i\Delta +\lambda_+)(i \Delta + \lambda_-)}
\end{equation}

\section{Device fabrication}
\label{appendix:b}

All devices were fabricated on $100\pm25~\mu$m thick C-plane (0001) Sapphire wafers with a diameter of $50.8~$mm. Wafers were cleaned with organic solvents (Toluene, Acetone, Methanol, Isopropanol, and DI water) in an ultrasonic bath to remove contamination, then were annealed at 1200$^\circ$C for 1.5 Hours. Prior to film deposition, wafers underwent a second clean with organic solvents (Toluene, Acetone, Methanol, Isopropanol, and DI water) in an unltrasonic bath, followed by 2 minute clean in 50$^\circ$C Nano-Strip\texttrademark{} etch, and a rinse with high purity DI water. The wafers then underwent a dehydration bake at 180$^\circ$C in atmosphere for 3 minutes. 

Immediately afterwards, wafers were loaded into an Ultratech Fiji G2 plasma enhanced atomic layer deposition tool for metallization, where they first underwent a 1 hour bake at 300$^\circ$C under vacuum continuously purged with $5~$sccm of argon gas. Chamber walls matched the substrate temperature. The deposition parameters and machine configuration are adapted from Ref. \cite{sowa2017peald}. ($t$-Butylimido)Tris(Diethylamido)-Niobium(V) (TBTDEN) was used as the niobium precursor, which was kept at 100$^\circ$C and delivered by a precursor Boost\texttrademark{} system, which introduces argon gas into the precursor cylinder to promote material transfer of the low vapor pressure precursor to  the wafer \cite{sowa2017peald}. The deposition cycle consisted of three 0.5 second pulses of boosted TBTDEN followed by 40 seconds of $300~$W plasma consisting of $80~$sccm hydrogen and $5~$sccm nitrogen. A flow of $5~$sccm of nitrogen and $10~$sccm of argon was maintained throughout the deposition process. After deposition the wafer was allowed to passively cool to 250$^\circ$C under vacuum. 

Following deposition, the wafers were cleaned with DI water in an ultrasonic bath to remove particulates, then underwent a dehydration bake at 180$^\circ$C in atmosphere for 3 minutes before spinning resist. For optical lithography, to avoid defocusing from wafer deformation, wafers were mounted to a Silicon handle wafer with AZ MiR 703 photoresist cured at 115$^\circ$C. Wafers were then coated with $\sim 1~\mu$m of positive I-line photoresist AZ MiR 703, and exposed with a Heidleberg MLA150 Direct Writer. For electron Beam lithography, wafers were first coated with $\sim 200~$nm of negative resist ARN 7520, followed by $40~$nm of conductive resist `Elektra' AR PC 5090, and then exposed with a Raith EBPG5000 Plus E-Beam Writer. The conductive coating was removed by a 60 second DI water quench. Both optical and E-Beam resists were baked at 110$^\circ$C to further harden the resist, and then developed for 60 seconds in AZ MIF 300, followed by a 60 second quench in DI water. We note that the rounded corners of our devices are by design to diffuse electric fields and reduce coupling to two level systems, and not defects induced by lithographic resolution.

The NbN films were etched in a Plasma-Therm inductively coupled plasma etcher. Etch chemistry, substrate etch depth and etch time have been shown to affect planar resonator quality factors \cite{lock2019surface}, in particular due to the formation of cross-linked polymers at the metal-resist interface after the bulk metal is etched away. For this reason we scale sample etch times to metal thickness, with a fixed over-etch time of 30 seconds to ensure complete metal removal. We use a Fluorine based ICP etch chemistry with a plasma consisting of $15~$sccm SF$_6$, $40~$sccm CHF$_3$, and $10~$sccm Ar. ICP and bias powers were kept at $100~$W, and the substrate was cooled to a temperature of 10$^\circ$C. Following etching, the resist was stripped in a combination of acetone and 80$^\circ$C Remover PG (N-Methyl-2-Pyrrolidone) which also serve to release the wafer from the Silicon carrier wafer. The wafers were then cleaned with organic solvents  (Acetone, Isopropanol, and DI water), coated with a $\sim 2~\mu$m protective layer of photoresist, and diced into $3.3 \times 2.3~$mm chips. These were stripped of protective resist with 80$^\circ$C Remover PG, cleaned with organic solvents  (Acetone, Isopropanol, and DI water), dried on an unpolished Sapphire carrier wafer in atmosphere at 80$^\circ$C, then mounted with Indium wire in the copper box described in the text.

\section{Film characterization}
\begin{figure}[htpb]
\centering
\includegraphics[width=300pt]{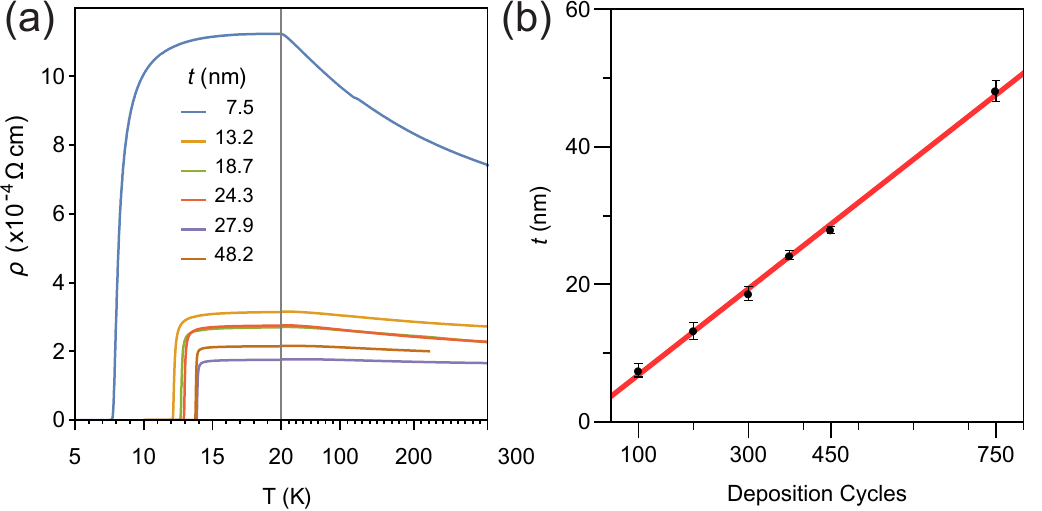}
\caption{Film measurements (a) Measured resistivity as a function of temperature showing decreasing resistivity with increasing temperature above the superconducting transition characteristic for NbN. (b) Thicknesses measured by profilometry as a function of deposition cycles, with a linear fit overlaid in red. We extract a growth rate of $0.63~$\AA\ per cycle.
\label{fig:figS2}}
\end{figure}

DC film characterization measurements were performed in a Quantum Design Physical Property Measurement System (PPMS) with a base temperature of $1.8~$K. Test structures consisting of a 1.5~mm$\times40~\mu$m wire were patterned on $7\times7~$mm chips going through the process described above along with device wafers, then wirebonded for four-wire measurements. Finished structures were kept in a low $\sim 500~$mTorr vacuum in an effort to minimize oxide growth prior to measurement, as we observed decreases up to $1~$K in critical temperatures for samples aged several days in atmosphere, likely a result of oxide growth \cite{medeiros2019measuringnbn} reducing the superconducting film thickness. 

After cooling the samples to $10~$K ($3~$K in the case of the $8~$nm film) in ambient magnetic fields, we verified that the residual resistance of the film dropped below the instrument noise floor of around $5\times10^{-3} \Omega$. After thermalizing for one hour, the samples were warmed up to $20~$K at a rate of 0.$1~$K/min, then warmed to $300~$K at a rate of $1~$K/min. In Fig. \ref{fig:figS2}(a) we plot measured resistivity as a function of temperature for the films in this study, which we use to extract $T_C$, $\rho_n$ and calculate $R_\Box$ and $L_\Box$ for the films. Similar to previous studies \cite{niepce2019nbnnanowirevap}, resistivity decreases with temperature above the superconducting transition, which is typical for strongly disordered materials \cite{mondal2011phasenbn,niepce2019nbnnanowirevap}.

\section{Millimeter-Wave Measurement Setup and calibration}
\begin{figure}[htbp]
\centering
\includegraphics[width=5.5in]{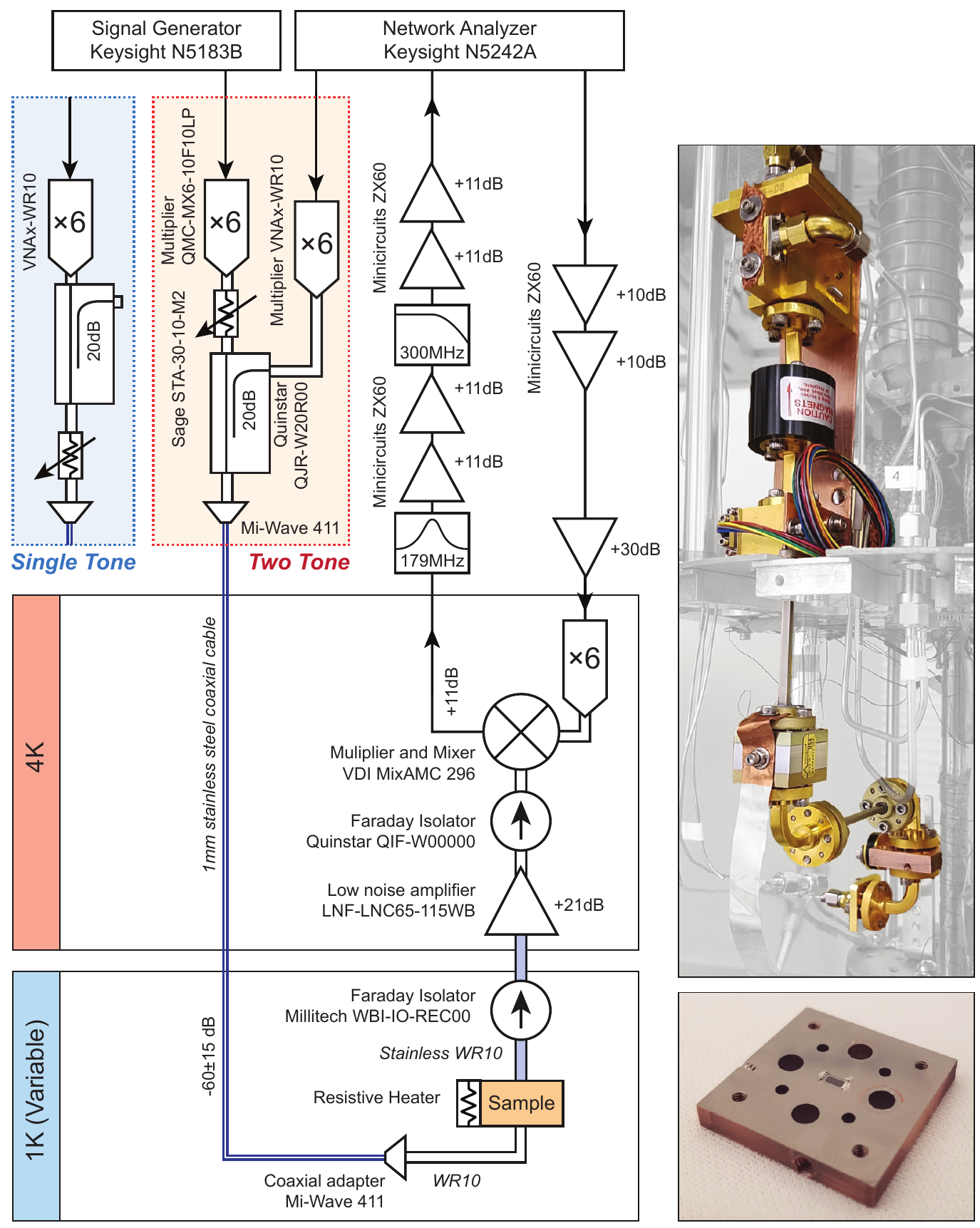}
\caption{Schematic of millimeter-wave measurement setup for single and two-tone configurations. Colored tabs show temperature stages inside the $^4$He adsorption refrigerator. A photograph on the right highlights relevant hardware inside the fridge. The bottom left shows a photograph of the sample with top waveguide removed. 
\label{fig:figS3}}
\end{figure}
All millimeter-wave characterization was performed in a custom built $^4$He adsorption refrigerator, with a base temperature of $0.9~$K, and a cycle duration of 3 hours. We generate millimeter-wave signals ($75$-$115~$GHz) at room temperature by sending microwave signals ($12$-$19~$GHz) into a frequency multiplier. We convert the generated waveguide TE$_{10}$ mode to a $1~$mm diameter stainless steel and beryllium copper coaxial cable, which carries the signal to the $1~$K stage of the fridge, thermalizing mechanically at each intermediate stage, then convert back to a WR-10 waveguide which leads to the device under test. The cables and waveguide-cable converters have a combined frequency-dependent loss ranging from $55.6~$dB to $75.8~$dB in the W-Band, which is dominated by the cable loss. We confirm the attenuation and incident device power at room temperature with a calibrated power meter (Agilent W8486A) and a referenced measurement with a VNAx805 receiver, however cryogenic shifts in cable transmission and minute shifts in waveguide alignment likely result in small variations in transmitted power. We are able to further confirm the applied power by measuring the lowest observed bifurcation point, and find that most bifurcation powers agree with predictions, yielding a maximum combined power uncertainty of approximately $\pm5~$dBm, which sets the uncertainty in our photon number measurements.

The sample is isolated from both millimeter-wave and thermal radiation from the $4~$K plate with two stainless steel waveguides 2 inches long and a faraday isolator. Using a resistive heater and a standard curve Ruthenium oxide thermometer we can perform temperature sweeps on the sample without significantly affecting the fridge stage temperatures. A low-noise amplifier ($T_N \sim 28~$K) amplifies the signal before passing through another faraday isolator, which further blocks any leaking signals. The signal then passes to a cryogenic mixer, which converts the signal to radio-wave ($100$-$300~$MHz) which we filter, amplify and measure at room temperature with a network analyzer. We control signal power by varying input attenuation and multiplier input power, confirming with room temperature calibrations as described above. For two-tone measurements, we move the signal path to the $20~$dB port of the input directional coupler, and add an additional frequency multiplier fed by a reference-locked microwave signal generator. For single-tone measurements, the $20~$dB port is capped with a short to minimize incident stray radiation.

\section{BCS conductivity temperature dependence}
\begin{figure}[b]
\centering
\includegraphics[width=6.69in]{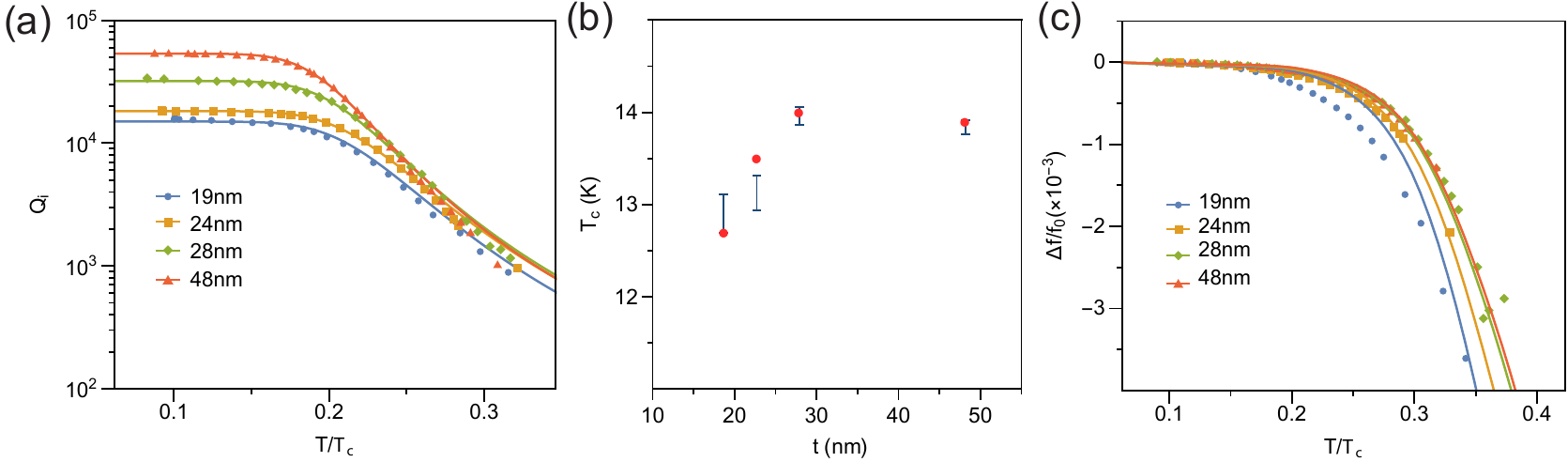}
\caption{Temperature dependence of BCS conductivity. (a) High power $Q_i$ as a function of normalized temperature for four resonators of different film thickness. Solid lines correspond to a BCS model with $T_c$ and kinetic inductance fraction $\alpha$ as fit parameters. (b) Extracted $T_c$ from fitting to BCS model (red dots), compared to $T_c$ from DC resistivity measurements. (c) Normalized frequency shift of the same resonators as a function of temperature, with overlaid predictions from the Mattis-Bardeen equations for $\sigma_2/\sigma_n$ with parameters taken from fits in (a). 
\label{fig:figS4}}
\end{figure}

Due to the large kinetic inductance fraction $\alpha$, or magnetic field participation ratio of the thin film resonators, we expect higher sensitivity to conductor loss, which in turn is sensitive to temperature. In Fig. \ref{fig:figS4}(a) we show the quality factor decrease as a function of temperature for resonators with four different thicknesses, with solid lines corresponding to a model of the form
\begin{equation}
    Q_i(T)^{-1}=Q_{i,\text{max}}^{-1} + Q_{\sigma}(T)^{-1}
\end{equation}
where $Q_{i,\text{max}}$ is a temperature independent upper bound arising from other sources of loss, and the conduction loss $Q_\sigma$ is given by \cite{reagor2016superconducting}:
\begin{equation}
    Q_\sigma(T)=\frac{1}{\alpha}\frac{\sigma_2(T,T_c)}{\sigma_1(T,T_c)}
\end{equation}
where $\sigma_1$ and $\sigma_2$ are the real and imaginary parts respectively of the complex surface impedance, calculated by numerically integrating the Mattis-Bardeen equations for $\sigma_1/\sigma_n$ and $\sigma_2/\sigma_n$ \cite{reagor2016superconducting,tinkham2004introduction,mattis1958theory}. We use $\alpha $ and $T_c$ as fit parameters in the model. Below $2~$K ( $T/T_c \sim 0.15$), $Q_i$ saturates, which indicates that conduction loss does not limit $Q_i$ for these devices. We note minor deviations from theory at higher temperatures, which may be a result of physical deviations from the standard curve calibrations used for the ruthenium oxide thermometer. Since these resonators were fabricated with $Q_e>10^4$ measuring resonators at higher temperatures where $Q_i$ is below $10^3$ proved experimentally challenging. In Fig. \ref{fig:figS4}(b) we plot the fitted $T_c$ values against those obtained with DC measurements and find reasonable agreement for higher thickness films, however note that the uncertainty in temperature calibration combined with the relatively low saturation values result in fitted $T_c$ uncertainties around $0.4~$K.

Bardeen-Cooper-Schrieffer theory also predicts a shift in London length as a function of temperature, which in the dirty (high disorder) limit is given by \cite{hazra2018microwave,tinkham2004introduction}:
\begin{equation}
    \frac{\lambda(T)}{\lambda(0)} = \frac{1}{\sqrt{\frac{\Delta(T)}{\Delta_0}\tanh\left(\frac{\Delta_0}{2k_bT}\right)}}
\end{equation}
We can measure this by tracking the resonant frequency shift. For sufficiently large kinetic inductance fractions, or $L_k \gg L_g$, the kinetic inductance will dominate the total inductance, so the normalized frequency shift will be \cite{hazra2018microwave}
\begin{equation}
\frac{f_0(T)}{f_0(0)} = \sqrt{\frac{\Delta(T)}{\Delta_0}\tanh\left(\frac{\Delta_0}{2k_bT}\right)}
\label{eq:fBCS}
\end{equation}
In Fig. \ref{fig:figS4}(c) shows the normalized frequency shift as a function of normalized temperature and predictions from Eq. \ref{eq:fBCS} with parameters $\alpha$ and $T_c$ taken from fits to $Q_i(T)$ above. Notably, we find significant deviation from the BCS theory for lower thicknesses, which has been previously observed for high-disorder films \cite{shearrow2018ald,beebe2016nbn,hazra2018microwave}.

\section{Controlling nonlinearity in the presence of additional losses}
\begin{figure}[b]
\centering
\includegraphics[width=6.69in]{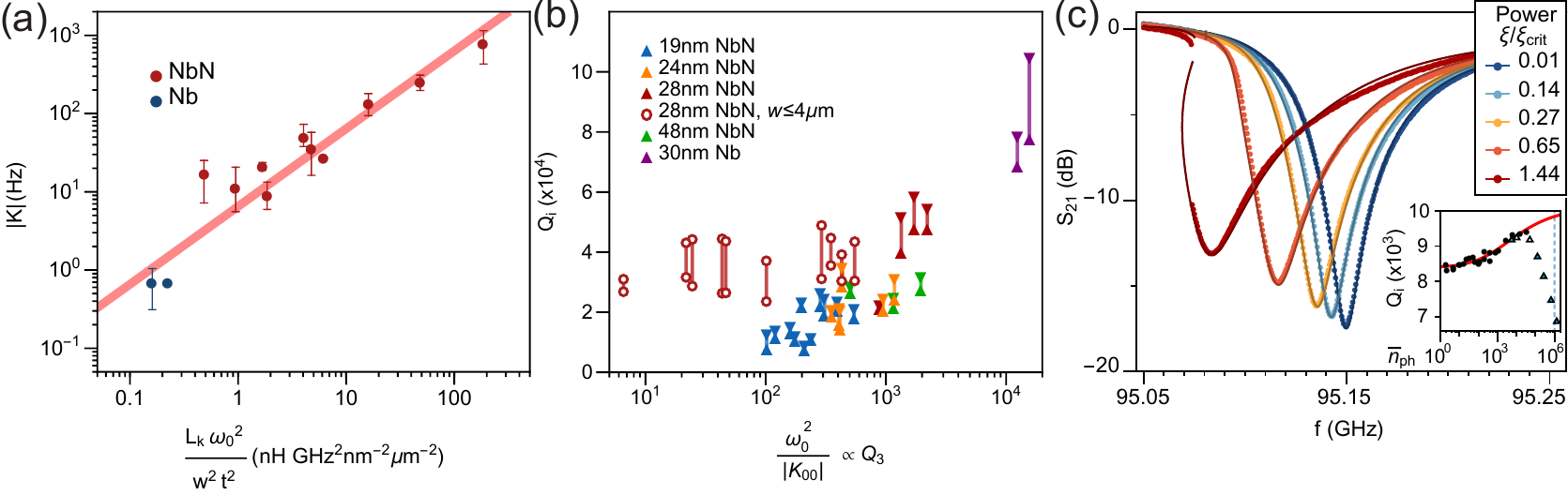}
\caption{(a) Self-Kerr constant $|K_{00}|$ as a function of parameters in Eq. \ref{eq:definekerr}, with a linear fit overlaid as a solid line. Points correspond to average values for groups of similar film thicknesses and wire widths, and bars represent the extrema of the group. (b) $Q_i$ as a function of $\omega_0^2/|K_{00}|$ which corresponds to the loss $Q_3$ associated with kinetic inductance. Points correspond to low and high power limits of $Q_i$. Note that devices with varying wire width (empty circles) do not appear to be correlated with $Q_3$. (c) Transmission as a function of frequency for a $18.7~$nm thick device at $95.15~$GHz taken at increasing powers $\xi$, with the inset highlighting decreasing $Q_i$ near the bifurcation power $\xi_\text{crit}$ (dashed blue line) deviating from two level system loss model (red line). Triangles correspond to nonlinear model fits, with traces shown in main panel marked in blue.
\label{fig:figS5}}
\end{figure}

From Ref. \cite{yurke2006performance}, we expect the self-Kerr nonlinearity originating from kinetic inductance of a $\lambda /4$ resonator to be
\begin{equation}
    K = -\frac{\hbar \omega_0^2}{I_c^2}\int_0^l dx u_0^4 \Delta L \propto -\frac{\hbar \omega_0^2 L_k}{J_c^2 w^2 t^2}
    \label{eq:definekerr}
\end{equation}
where in our case the nonlinear kinetic inductance $\Delta L$ is constant along the transmission line, so integrating over the fundamental mode profile $u_0$ yields a constant. We have also transformed the critical current $I_c$ into a critical current density $J_c$, and used the assumption that the nonlinear kinetic inductance is proportional to the linear kinetic inductance \cite{kher2017superconducting,swenson2013operation}. Fig. \ref{fig:figS5}(a) expands on Fig. 3(b), showing measured self-Kerr constants for all resonators in this study (grouped into points by film thickness and wire width) as a function the parameters in Eq. \ref{eq:definekerr}, with the solid line corresponding to a linear fit. We have also added data from identical resonators fabricated from 30nm electron-beam evaporated niobium to extend the parameter range. We find reasonable agreement with dependence on the parameters in Eq. \ref{eq:definekerr}, however note that the dependence is much less clear than that on wire width $w$.

Nonlinear kinetic inductance is also associated with a nonlinear resistance of the same form $R = R_0 + \Delta R I^2/I_c^2$. Based on Ref. \cite{yurke2006performance}, and assuming the nonlinear resistance scales with kinetic inductance, losses associated with nonlinear resistance will be
\begin{equation}
    \gamma_3 = \frac{\omega_0}{Q_3} = \frac{3\hbar \omega_0}{8I_c^2}\int_0^l dx u_0^4 \Delta R
    \propto \frac{\hbar \omega_0 L_k}{J_c^2 w^2 t^2} \sim \frac{-K}{\omega_0}
\end{equation}
This indicates that upper bounds on nonlinear losses should scale as $Q_3 \sim \frac{\omega_0^2}{|K|}$. In Fig. \ref{fig:figS5}(b) we plot low and high power limits of $Q_i$ devices in this study with the addition of $30~$nm Niobium devices described above, and find that for resonators with fixed wire widths $w=4~$$\mu$m, there appears to be a potential correlation of $Q_i$ with $Q_3$ indicating nonlinear resistance may be a potential loss mechanism. 

In our analysis, we have also neglected to take into account higher harmonics of the resonator, which will be coupled to the fundamental mode by cross-Kerr interactions $\chi_{mn}$, which for evenly spaced harmonics scale as \cite{yurke2006performance}
\begin{equation}
    \chi_{mn} = -\frac{3\hbar \omega_m \omega_n}{I_c^2}\int_0^l dx u_m^2u_n^2 \Delta L
    \propto K
\end{equation}
Given the proportionality to $K$, the correlation described above may also potentially be a result of cross-Kerr effects. For line-widths large enough to cover any deviations from evenly spaced higher harmonics, we anticipate see power dependent conversion processes: in particular for a Kerr nonlinear system with harmonics spaced at $\omega_0$ and $3\omega_0$, at powers approaching the critical power we would expect increased conversion efficiency from the fundamental to third harmonic \cite{hashemi2009nonlinear}, which in our experiment would be observed as increased resonator loss at higher powers. 

In Fig. \ref{fig:figS5}(c) we show the atypical transmission spectra of a $18.7~$nm thick, $4~$$\mu$m wide device with a particularly large line-width showing decreasing $Q_i$ near the bifurcation power (above $n_\text{ph} \sim 10^5$), departing from the two-level system model described in the main text. This additional power-dependent loss may be the result of the nonlinear mechanisms described above, but may also be a result of circulating currents exceeding the thin film critical current density, which is lowered by the increased London lengths of the thinner films \cite{talantsev2015universal,tsukamoto2005ac}. However since the loss could also simply be a result of frequency dependent dissipation, the source remains unclear.

In Fig. \ref{fig:figS5}(b), we also observe that resonators achieving higher nonlinearities by reducing wire width do not appear to be affected by the nonlinear loss rate described above. We also find that these devices do not showcase obvious signs of high power loss shown in Fig. \ref{fig:figS5}(c). While this may be a result of the difference in fabrication methods (see Appendix \ref{appendix:b}), the thinner wires may have higher vortex critical fields \cite{stan2004critical} leading to reduced vortex formation, and thus lower loss associated with vortex dissipation. Additionally, the thinner wires at the shorted end of the quarter wave section of the resonator further shift the higher harmonics, potentially resulting in lower cross-Kerr conversion loss.

\end{document}